\documentclass{iopart}
\def\RR{{\rm I\kern-.1567em R}}                              % Doppel R
 \def\CC{{\rm C\kern-4.7pt                                    % Doppel C
 \vrule height 7.7pt width 0.4pt depth -0.5pt \phantom {.}}}
 \def\ZZ{{\sf Z\kern-4.5pt Z}}
\usepackage{graphics, graphicx}

\begin{document}

\title[k-defects as compactons]
{$k$-defects as compactons }

\author{C. Adam$^1$, J.
S\'{a}nchez-Guill\'{e}n$^1$ and A. Wereszczy\'{n}ski$^2$}

\address{$^1$ Departamento de Fisica de Particulas, Universidad
       de Santiago and Instituto Galego de Fisica de Altas Enerxias
       (IGFAE) E-15782 Santiago de Compostela, Spain}

\address{$^2$ Institute of Physics,  Jagiellonian University,
       Reymonta 4, 30-059 Krak\'{o}w, Poland}

\eads{ \mailto{adam@fpaxp1.usc.es} \mailto{joaquin@fpaxp1.usc.es}
\mailto{wereszczynski@th.if.uj.edu.pl}}

\begin{abstract}
We argue that topological compactons (solitons with compact
support) may be quite common objects if $k$-fields, i.e., fields
with nonstandard kinetic term, are considered, by showing
that even for models with well-behaved potentials the unusual
kinetic part may lead to a power-like approach to the vacuum, which is a
typical signal for the existence of compactons. The related
approximate scaling symmetry as well as the existence of
self-similar solutions are also discussed.
\\
As an example, we discuss domain walls in a potential Skyrme model
with an additional quartic term, which is just the standard
quadratic term to the power two. We show that in the critical
case, when the quadratic term is neglected, we get the so-called
quartic $\phi^4$ model, and the corresponding topological defect
becomes a compacton. Similarly, the quartic sine-Gordon compacton
is also derived.
\\
Finally, we establish the existence of topological half-compactons and
study their properties.
\end{abstract}
%%%%%%%%%%%%%%%%%%%%%%%%%%%%%%%%%%%%%%%%%%%%%%%%%%%%%%%%%%%%%%
\section{Introduction}
%%%%%%%%%%%%%%%%%%%%%%%%%%%%%%%%%%%%%%%%%%%%%%%%%%%%%%%%%%%%%
The present paper  investigates  the connection
between $k$-fields, i.e., fields with dynamics governed by a
nonstandard kinetic term, and the appearance of a very special class of
topological defects called {\it compactons}, that is, solitons
which approach the vacuum value at a finite distance.
\\
Classical field theories with a nonstandard kinetic term or,
more generally, gradient term find more and more
interesting applications in various branches of modern
theoretical physics. Originally, such unusual gradients have
been included to stabilize static solutions in some soliton models
as, e.g., in the Skyrme \cite{skyrme1} or Faddeev-Niemi models
\cite{fn}. There the additional quartic term scales oppositely to the
quadratic kinetic energy, such that the Derrick scaling argument
against the existence of static finite-energy solutions is circumvented. A
different strategy to circumvent  Derrick's argument
led to the investigation of
Lagrangians where instead of the quadratic term one has a
fractional power of the gradient. Moreover, due to the scaling
symmetry, such models allow for the analytic calculation of
exact chiral solitons
\cite{deser}, \cite{we1} or knot solitons
\cite{nicole}-\cite{we2}.
\\
Quite recently, models with generalized dynamics, i.e., with
$k$-fields, have become particularly popular in cosmology. They
have been proposed in the context of inflation leading to
$k$-inflation, i.e., $k$-essence models \cite{ap1}-\cite{ap2}, or as
an alternative solution to the problem of dark matter
\cite{dark1}, \cite{dark2} (see, e.g., MOND, that is, modified
Newtonian dynamics models \cite{mond}).
\\
The behavior of topological defects (domain walls, vortices and
monopoles) in models with a $k$-deformed kinetic part have also been
studied \cite{babichev}, \cite{bazeia}. In general, it has
been observed that the influence of the non-quadratic kinetic term
leads to quantitative rather than qualitative differences.
As we will show here, however, a deformation of the kinetic
part can result in a more profound change of the properties of
topological defects, namely, in the existence of compactons.
\\
As is suggested by its name, a compacton is a soliton with compact
support. Compactons have originally been discovered as a special class
of solitary waves in generalized versions of the KdV equation
\cite{rosenau1}-\cite{cooper3}. Further examples of compactons as,
for instance, breather-like solutions are also known
\cite{rosenau3}, \cite{dinda}.
\\
Recently, compactons have been discussed in topologically
nontrivial systems, as well. Such a topological compacton is an
object which reaches the exact vacuum value at a finite distance.
This class of topological defects naturally occurs in models with
the standard quadratic kinetic scalar term provided that so-called
$V$-shaped potentials (i.e., potentials which are not smooth at
their minimum \cite{arodz1}-\cite{arodz5}) are considered. In
particular, the left and right derivatives at the minimum do not
vanish and the second derivative does not exist. In practice, this
means that there is no mass scale in the system.
\\
The most striking features characterizing compactons in standard
models with $V$-shaped potentials are the parabolic approach to
the vacuum value and the existence of an approximate scaling
symmetry. Both effects are universal and do not depend on the
particular form of the potential.
\\
Interestingly, some $V$-shaped models can describe the pinning
of topological defects to a boundary or an impurity.
\\
Here, we show that compactons can emerge due to the non-standard
kinetic term even when the potential is assumed to be an
analytical function of the scalar field $\xi$. In general, the
appearance of compactons is a  result of the mutual relation
between the kinetic (more precisely spatial gradient term) and
the potential part of the action.
\\
The paper is organized as follows. In the next section we give a
general overview of the Bogomolny sector for $(1+1)$ dimensional field
theories with non-standard kinetic term. In Section 3 the general
relation between compactons and $k$-defects in $(1+1)$ dimensions
is established.
In particular, we find that the powerlike approach to the vacuum and the
approximate scaling symmetry are present in our case, as well.
Section 4 is devoted to the discussion of domain wall type
compactons in the example of a scalar field theory embedded
into a modified Skyrme model. Also the issue of stability is discussed in some
detail. 
In Section 5 we show that for slightly more general potentials half-compactons
may exist. Half-compactons approach different vacuum values in a different
manner, approaching some vacua at a finite distance,
whereas they show the usual kink behaviour
(exponential approach) for other vacua.
Section 6 contains our conclusions.
%%%%%%%%%%%%%%%%%%%%%%%%%%%%%%%%%%%%%%%%%%%%%%%%%%%%%%%%%%%%%%
\section{Bogomolny equation for non-quadratic Lagrangians in
(1+1) dimensions}
%%%%%%%%%%%%%%%%%%%%%%%%%%%%%%%%%%%%%%%%%%%%%%%%%%%%%%%%%%%%%%
It is a well known fact that for any scalar model in (1+1)
dimensions with the usual quadratic kinetic term and arbitrary
potential, there exists a Bogomolny sector defined by a certain
first order differential equation. Of course, in order to speak
about the Bogomolny sector a nontrivial topology must exist.
Therefore, we assume that there are at least two distinct ground
states for the scalar field. The existence of such a first order
equation results in the following properties \cite{maison}:
\\
{\it i)} the second order Euler-Lagrange equation of motion is
identically satisfied by solutions of the Bogomolny equation.
\\
{\it ii)} the energy of the Bogomolny solution is completely
determined by its topology.
\\
{\it iii)} the static energy-stress tensor vanishes identically.
\\
In fact, these properties of the Bogomolny solutions are valid also
in other, higher dimensional cases (2 dimensional vortices in the
Abelian Higgs model, 't Hooft-polyakov monopoles in the
non-Abelian Higgs model, instantons of the 4-dimensional
Yang-Mills theory).
\\
Of course, the models investigated here have a more complicated
kinetic part. Therefore, it seems reasonable to look at this class of
Lagrangians from a slightly more general point of view
before analyzing domain walls in more specific models.
First of all, it is important to understand the Bogomolny
solution in these models.
\\
Here, we show that one can easily find the Bogomolny equation for
non-quadratic models. Additionally, we prove that the
 three properties mentioned above still hold.
%%%%%%%%%%%%%%%%%%%%%%%%%%%%%%%%%%%%%%%%%%%%%%%%%%%%%%%%%%%%%%
\subsection{Solution of the Bogomolny equation solves the equation of motion}
%%%%%%%%%%%%%%%%%%%%%%%%%%%%%%%%%%%%%%%%%%%%%%%%%%%%%%%%%%%%%%
Let us consider the most general form of a Lorentz invariant
Lagrangian depending on the scalar field and its first derivatives
\begin{equation}
L=L(v,\xi), \label{model bog}
\end{equation}
where
\begin{equation}
v \equiv \frac{1}{2} \partial_{\mu} \xi \partial^{\mu} \xi
\label{variable}
\end{equation}
(such models have been investigated, e.g., in \cite{bazeia}, \cite{AlGa}).
The equation of motion reads
\begin{equation}
\partial_{\mu} \left( L_v \partial^{\mu} \xi \right) -
L_{\xi}=0. \label{gen motion}
\end{equation}
For static configurations it can be rewritten as
\begin{equation}
 \left( L_w  \xi_x \right)_x + L_{\xi}=0,
\label{gen static motion}
\end{equation}
where now
\begin{equation}
w \equiv -\frac{1}{2}\xi_x^2
\end{equation}
and $L$ is the static part of the Lagrangian.
This equation can be integrated to
\begin{equation}
L+L_w \xi_x^2=0 \label{gen bogom}
\end{equation}
where the integration constant has been set to zero, which defines the
Bogomolny sector in the one-dimensional case.
By construction, all solutions of the first order equation
(\ref{gen bogom}) satisfy the second order field equation, as well.
Moreover, due to the Lorentz invariance of the model we obtain a
travelling (boosted) solution
\begin{equation}
\xi (x,t)=\tilde \xi (\gamma(x\pm \beta t)),
\end{equation}
where $\tilde \xi$ is a solution of (\ref{gen bogom}) and $|\beta|<1$.
%%%%%%%%%%%%%%%%%%%%%%%%%%%%%%%%%%%%%%%%%%%%%%%%%%%%%%%%%%%%%%
\subsection{Space component of the static stress tensor vanishes
identically}
%%%%%%%%%%%%%%%%%%%%%%%%%%%%%%%%%%%%%%%%%%%%%%%%%%%%%%%%%%%%%%
The corresponding energy-stress tensor has the form
\begin{equation}
T^{\mu \nu}=L_v\partial^{\mu}\xi \partial^{\nu} \xi - g^{\mu \nu}
L. \label{energy stress}
\end{equation}
In particular, one finds that
\begin{equation}
T^{01}=L_v \xi_t \xi_x,
\end{equation}
\begin{equation}
T^{11}=L_v\xi_x^2+L.
\end{equation}
Of course, in the static case the component $T^{01} =0$.
Additionally, in the Bogomolny sector
\begin{equation}
T^{11}=L_w\xi_x^2+L \equiv 0 .
\end{equation}
Therefore, for solutions of the Bogomolny equation the
space-like component of the static energy-stress tensor vanishes
identically.
%%%%%%%%%%%%%%%%%%%%%%%%%%%%%%%%%%%%%%%%%%%%%%%%%%%%%%%%%%%%%%
\subsection{The energy of the solution depends only on its topology}
%%%%%%%%%%%%%%%%%%%%%%%%%%%%%%%%%%%%%%%%%%%%%%%%%%%%%%%%%%%%%
The remaining non-vanishing component of the tensor gives the
energy of the system. Here we have
\begin{equation}
E=\int_{-\infty}^{\infty} T^{00}dx=\int_{-\infty}^{\infty}
(L_v\xi^2_t - L) dx.
\end{equation}
In the Bogomolny sector,
\begin{equation}
E=\int_{-\infty}^{\infty} L_w \xi_x^2 dx .
\end{equation}
On the other hand, equation (\ref{gen bogom}) can be viewed as a
complicated equation for $\xi_x$. A formal solution reads
\begin{equation}
\xi_x=\mathcal{F} (\xi).
\end{equation}
Then,
\begin{equation}
\fl E= \int_{-\infty}^{\infty} L_w \mid_{\xi_x=\mathcal{F} (\xi)}
\mathcal{F} (\xi) \xi_x dx= \int_{\xi(-\infty)}^{\xi(\infty)} L_w
\mid_{\xi_x=\mathcal{F} (\xi)} \mathcal{F} (\xi) d\xi.
\end{equation}
In other words, for solutions of the Bogomolny equation, the
energy depends only on the boundary conditions for the scalar field,
i.e., on the topology of the solution.
\\ \\
It is easy to notice that this observation is true only if the
scalar field reaches its vacuum value at infinity. Indeed, then
the last equality holds and the energy is fixed by topology. In the
opposite case, when the vacuum is approached at a finite distance, the
situation changes. The global topology does no longer
define the energy of the system. Instead, it is possible
for a fixed total topological charge to construct solutions with
various energies describing  different collections of compactons and
anti-compactons.
\\
Therefore, compactons give a first example of Bogomolny-type
solutions for which there is no one-to-one correspondence between
global topology and energy.
%%%%%%%%%%%%%%%%%%%%%%%%%%%%%%%%%%%%%%%%%%%%%%%%%%%%%%%%%%%%%%
\section{k-fields and compactons}
%%%%%%%%%%%%%%%%%%%%%%%%%%%%%%%%%%%%%%%%%%%%%%%%%%%%%%%%%%%%%%
In the subsequent investigation we consider $k$-deformed Lorentz
invariant Lagrangians. Nonetheless, as we analyze mainly the
static configurations, our compactons may be solutions to other
models, provided they have the same form in the static regime.
Therefore they can also exist in some effective (non-Lorentz
invariant) models, where the time evolution is governed by the
usual second time derivative \cite{dusual}.
%%%%%%%%%%%%%%%%%%%%%%%%%%%%%%%%%%%%%%%%%%%%%%%%%%%%%%%%%%%%%%
\subsection{Quartic models}
%%%%%%%%%%%%%%%%%%%%%%%%%%%%%%%%%%%%%%%%%%%%%%%%%%%%%%%%%%%%%%
We begin our analysis of the compactons in $k$-deformed
Lagrangians with the {\it quartic} model
\begin{equation}
L=|\xi_{\nu}\xi^{\nu}| \xi_{\mu}\xi^{\mu} - U, \label{quartic
model}
\end{equation}
where instead of the standard quadratic kinetic term one deals
with its quartic version. The potential $U$ is assumed to be a
smooth function of the field $\xi$ with vanishing derivative at the
local minimum.
\\
The scalar field satisfies the following equation of motion
\begin{equation}
4 \partial_{\mu} [|\xi_{\nu}\xi^{\nu}| \xi^{\mu}] + U_{\xi}=0.
\end{equation}
%%%%%%%%%%%%%%%%%%%%%%%%%%%%%%%%%%%%%%%%%%%%%%%%%%%%%%%%%%%%%%
\subsection{Parabolic approach}
%%%%%%%%%%%%%%%%%%%%%%%%%%%%%%%%%%%%%%%%%%%%%%%%%%%%%%%%%%%%%%
For the static configurations we easily derive the Bogomolny first
order equation
\begin{equation}
3\xi^4_{x}=U. \label{quatic mod bog}
\end{equation}
In the vicinity of the local minimum located at $\xi_0$ the
potential can be expanded into the Taylor series
\begin{equation}
U=U (\xi_0) + U'(\xi_0) (\xi-\xi_0) +\frac{1}{2}U''(\xi_0)
(\xi-\xi_0)^2+...
\end{equation}
From the smoothness of the potential we get $U'(\xi_0)=0$. Then,
in the neighborhood of the minimum, a small perturbation $\delta
\xi$ of the vacuum value of the scalar field
\begin{equation}
\xi=\xi_0+\delta \xi, \label{def pert}
\end{equation}
obeys
\begin{equation}
(\delta \xi_x)^4 = \alpha (\delta \xi)^2,
\end{equation}
where $\alpha^2= U''(\xi_0)/6$. Here we consider the case with
non-vanishing $U''(\xi_0)$. The obvious solution reads
\begin{equation}
\delta \xi=\left( \frac{\alpha}{4} \right)^2 x^2.
\end{equation}
Therefore, in the quartic model the vacuum state is approached parabolically.
%%%%%%%%%%%%%%%%%%%%%%%%%%%%%%%%%%%%%%%%%%%%%%%%%%%%%%%%%%%%%%
\subsection{Scaling symmetry}
%%%%%%%%%%%%%%%%%%%%%%%%%%%%%%%%%%%%%%%%%%%%%%%%%%%%%%%%%%%%%%
Let us consider the full dynamical field equation near the local
minimum
\begin{equation}
4 \partial_{\mu} [|\xi_{\nu}\xi^{\nu}| \xi^{\mu}] +
U''(\xi_0) (\xi-\xi_0) =0. \label{quatic mod symmetry}
\end{equation}
One can check that if the field $\delta \xi$ obeys this equation
then
\begin{equation}
\delta \xi_{\lambda}=\lambda^2 \delta \xi \left(
\frac{x}{\lambda}, \frac{t}{\lambda} \right)
\end{equation}
also is a solution of (\ref{quatic mod symmetry}). This scaling
symmetry is identical to the symmetry originally found in the
$V$-shaped models. The symmetry is an approximate one as it exists
only in an infinitesimal neighbourhood of the local minimum.
\\
In the case of the parabolic potential, such a scaling symmetry
becomes exact. Then, in this {\it quartic harmonic
oscillator} model one can construct self-similar solutions
\begin{equation}
\xi (x,t)= x^2 W \left( \frac{t}{x} \right).
\end{equation}
This model can be viewed as the quartic counterpart of the
signum-Gordon model discussed by Arodz.
\\
As we see, the quartic model shares the universal features of the
standard quadratic model with $V$-shaped potentials. However,
while the standard compactons solve the field equations for all
$x$ except the point where the vacuum value is reached (they are
solutions in the weak sense), the compactons in the quartic model
obey the field equations everywhere. This is true in spite of the fact
that the second derivative of the scalar field in general is not
smooth at the boundary of the compacton. It is given by a step
function. However, in the field equation of the quartic model this
second derivative is multiplied by first derivatives, and the
first derivative generically vanishes at the boundary of the
compacton, rendering the field equation well-defined even there.
%%%%%%%%%%%%%%%%%%%%%%%%%%%%%%%%%%%%%%%%%%%%%%%%%%%%%%%%%%%%%%
\subsection{Generalization}
%%%%%%%%%%%%%%%%%%%%%%%%%%%%%%%%%%%%%%%%%%%%%%%%%%%%%%%%%%%%%%
A more general approach  to the vacuum value, but still at
finite distance, is realized in the following model
\begin{equation}
L=|\xi_{\nu}\xi^{\nu}|^n \xi_{\mu}\xi^{\mu} - U, \label{n model}
\end{equation}
where $n>-1/2$ and $n \neq 0$. This family of the Lagrangians
belongs to the admissible theories considered by \cite{AlGa}.
\\
In fact, near the local minimum located at $\xi_0$ we get
\begin{equation}
\xi=\xi_0 + \frac{n}{1+n}\left(
\frac{n^2U''(\xi_0)}{2(n+1)^2(2n+1)} \right)^{\frac{1}{2n}}
x^{\frac{n+1}{n}}.
\end{equation}
Similarly, the corresponding approximate scaling symmetry is
modified and takes the form
\begin{equation}
\delta \xi_{\lambda}=\lambda^{1+n} \delta \xi \left(
\frac{x}{\lambda}, \frac{t}{\lambda} \right).
\end{equation}
Again, this symmetry is  exact if the potential takes the
form
\begin{equation}
U=|\xi|^{n+1}.
\end{equation}
The pertinent self-similar solutions can be found using the Ansatz
\begin{equation}
\xi (x,t)= x^{1+n} W \left( \frac{t}{x} \right).
\end{equation}
Let us finally notice that compactons should exist for all
possible Lagrangians which asymptotically, for small values of
$v=(1/2) \xi_{\mu}^2$, take the form given by expression (\ref{n
model}). Then, the scalar field approaches the vacuum in a
power-like manner and no exponential tail exists.
%%%%%%%%%%%%%%%%%%%%%%%%%%%%%%%%%%%%%%%%%%%%%%%%%%%%%%%%%%%%%%
\section{Example: Skyrme model with new quartic term}
%%%%%%%%%%%%%%%%%%%%%%%%%%%%%%%%%%%%%%%%%%%%%%%%%%%%%%%%%%%%%%
%%%%%%%%%%%%%%%%%%%%%%%%%%%%%%%%%%%%%%%%%%%%%%%%%%%%%%%%%%%%%%
\subsection{Model}
%%%%%%%%%%%%%%%%%%%%%%%%%%%%%%%%%%%%%%%%%%%%%%%%%%%%%%%%%%%%%%
As a particular example of a theory with unusual kinetic term we
use the Skyrme model modified by a new quartic part. Namely,
\begin{equation}
\mathcal{L}=\frac{m^2}{2} \mathcal{L}_2 - M^2 \mathcal{L}_4
+\tilde{M}^2 \tilde{\mathcal{L}}_4- \mathcal{L}_0, \label{model}
\end{equation}
where
\begin{equation}
\mathcal{L}_2= tr (U^{\dagger} \partial_{\mu} U U^{\dagger}
\partial^{\mu} U) \label{quadratic}
\end{equation}
and
\begin{equation}
\mathcal{L}_4= tr [U^{\dagger} \partial_{\mu} U, U^{\dagger}
\partial_{\nu} U]^2. \label{quartic}
\end{equation}
As in the standard Skyrme model, $U$ is a SU(2) valued field living
in $(3+1)$ dimensional space-time. Moreover, $m, M$ and
$\tilde{M}$ are parameters. The new quartic term is chosen in the
following form
\begin{equation}
\tilde{\mathcal{L}}_4= |\mathcal{L}_2| \mathcal{L}_2.
\end{equation}
In addition, we include a potential term $\mathcal{L}_0$ which will
be specified below.
\\
The matrix field can be parameterized by
\begin{equation}
U=e^{i\vec{\xi} \vec{\sigma}},
\end{equation}
where $\vec{\sigma}$ are the Pauli matrices and $\vec{\xi}$ is a
three component real vector field. However, it is convenient to
adopt a different parametrization
\begin{equation}
U=e^{i\xi \vec{n} \vec{\sigma}}, \label{def u}
\end{equation}
where
\begin{equation}
\xi=|\vec{\xi}|, \;\;\; \vec{n}=\frac{\vec{\xi}}{|\vec{\xi}|}
\label{def xi}
\end{equation}
and the unit vector field is expressed by a complex scalar field $u$
via stereographic projection
\begin{equation}
\vec{n}= \frac{1}{1+|u|^2} \left( u+u^*, -i(u-u^*), |u|^2-1
\right). \label{stereo}
\end{equation}
In terms of the new variables we find that
\begin{equation}
\mathcal{L}_2=\xi_{\mu}\xi^{\mu}+4\sin^2 \xi
\frac{u_{\mu}\bar{u}^{\mu}}{(1+|u|^2)^2} \label{quadratic_new}
\end{equation}
\begin{equation}
\fl \mathcal{L}_4=16 \sin^2 \xi \left(
\xi_{\mu}\xi^{\mu}\frac{u_{\mu}\bar{u}^{\mu}}{(1+|u|^2)^2} -
\frac{\xi^{\mu} u_{\mu} \xi_{\nu} \bar{u}^{\nu}}{(1+|u|^2)^2}
\right) +16 \sin^4 \xi \frac{(u_{\mu}\bar{u}^{\mu})^2
-u_{\mu}^2\bar{u}_{\nu}^2}{(1+|u|^2)^4}. \label{quartic_new}
\end{equation}
In order to consider domain walls, we assume that
the complex field $u$ is trivial, i.e.,
it takes its vacuum value everywhere,
\begin{equation}
u=0 \;\;\; \Rightarrow \;\;\; \vec{n}=(0,0,1). \label{ansatz}
\end{equation}
As a consequence, we deal with the following Lagrangian
\begin{equation}
L=\frac{m^2}{2} \xi_{\mu}\xi^{\mu} + \tilde{M}^2
|\xi_{\mu}\xi^{\mu}|\xi_{\mu}\xi^{\mu} - L_0 (\xi). \label{model
full domain}
\end{equation}
As we see, the appearance of the new quartic term qualitatively
changes the Lagrangian describing the dynamics of domain walls. In
contrast to the usual Skyrme model where the quartic Skyrme
term identically vanishes if Ansatz (\ref{ansatz}) is assumed, the
new quartic part does contribute to the Lagrangian.
\\
In the subsequent analysis we consider two particular potentials,
which in our parametrization are the
well-known  $\phi^4$ potential
\begin{equation}
L_0=3\lambda^2 (\xi^2-a^2)^2
\end{equation}
or the sine-Gordon potential
\begin{equation}
L_0=\frac{3}{2} \lambda^2 (1-\cos \xi) ,
\end{equation}
respectively. Here, $\lambda$ is a real constant. Moreover, we
restrict the space-time to (1+1) dimensions. From the point of
view of the modified Skyrme model it means that physical
quantities like energy density or energy are given per unit area.
\\
Notice that this model slightly differs from the class of
Lagrangians which have been recently analyzed in \cite{babichev},
\cite{bazeia}. In fact, the model considered by these authors
reads
\begin{equation}
L=\frac{m^2}{2} \xi_{\mu}\xi^{\mu} - \tilde{M}^2
(\xi_{\mu}\xi^{\mu})^2 - L_0 (\xi) \label{model full domain
bazeia}
\end{equation}
and corresponds to a modified new quartic term
$\tilde{\mathcal{L}}_4=-\mathcal{L}_2^2$. In spite of the fact
that both Lagrangians (\ref{model full domain}), (\ref{model full
domain bazeia}) lead to the same static solutions, the
time-dependent configurations behave differently. In fact, the
energy of the second model is not bounded from below, as was
already observed in Ref. \cite{babichev}. We comment briefly on
this point in Subsection 4.5.
%%%%%%%%%%%%%%%%%%%%%%%%%%%%%%%%%%%%%%%%%%%%%%%%%%%%%%%%%%%%%%
\subsection{$\tilde{M}=0$ i.e. $\phi^4$ domain walls}
%%%%%%%%%%%%%%%%%%%%%%%%%%%%%%%%%%%%%%%%%%%%%%%%%%%%%%%%%%%%%
Let us begin our discussion with two relatively simple cases
corresponding to two rather special values of the parameters,
namely $\tilde{M}=0$ or $m=0$.
\\
The first possibility means that the new quartic term is absent and we
get the standard case with the $\phi^4$ potential. Of
course, the corresponding domain wall solution is just the $\phi^4$
kink
\begin{equation}
\xi (x) = a \tanh \left[ \frac{\sqrt{6} a\lambda}{m} (x+x_0)
 \right]. \label{phi4 kink}
\end{equation}
%%%%%%%%%%%%%%%%%%%%%%%%%%%%%%%%%%%%%%%%%%%%%%%%%%%%%%%%%%%%%%
\subsection{$m=0 $ i.e. the quartic $\phi^4$ compactons}
%%%%%%%%%%%%%%%%%%%%%%%%%%%%%%%%%%%%%%%%%%%%%%%%%%%%%%%%%%%%%
The second special case is more interesting. It is obtained
when the parameter $m$ vanishes, $m=0$, that is,
we neglect the standard, quadratic kinetic
part of the model. Then we arrive at the following Lagrangian
\begin{equation}
L=\tilde{M}^2 |\xi_{\mu}\xi^{\mu}|\xi_{\mu}\xi^{\mu} - 3\lambda^2
(\xi^2-a^2)^2, \label{model compacton}
\end{equation}
In accordance with the previous Section we call this model the {\it
quartic $\phi^4$ model}. The pertinent Bogomolny equation
reads
\begin{equation}
\xi^4_x=\frac{\lambda^2}{ \tilde{M}^2 }(\xi^2-a^2)^2. \label{bogom
comp}
\end{equation}
It is worth underlining that the same Bogomolny equation can be
derived for the model (\ref{model full domain bazeia}) provided $m=0$,
i.e.,
\begin{equation}
L=-\tilde{M}^2 (\xi_{\mu}\xi^{\mu})^2 - 3\lambda^2 (\xi^2-a^2)^2,
\label{model compacton bazeia}
\end{equation}
It is obvious that the static sectors of both models are
identical. One can solve this equation and get the single
compacton solution
\begin{equation}
\xi (x) = \left\{
\begin{array}{lc}
- a & x \leq - \frac{\pi}{2} \sqrt{\frac{\tilde{M}}{\lambda}} \\
a \sin \sqrt{\frac{\lambda}{\tilde{M}}}x &
-\frac{\pi}{2}\sqrt{\frac{\tilde{M}}{\lambda}} \leq x \leq
\frac{\pi}{2}\sqrt{\frac{\tilde{M}}{\lambda}}  \\
a & x \geq \frac{\pi}{2}\sqrt{\frac{\tilde{M}}{\lambda}},
\end{array}
\right. \label{compacton sol}
\end{equation}
which interpolates between the two distinct vacuum values $-a$ and
$a$.\footnote{Such a compacton solution has been originally found in a
non-relativistic model describing a mechanical system of coupled
pendulums \cite{dusual}. It follows from the fact that the static
regimes of the model representing a continuous idealization of the
system and the quartic sine-Gordon model are identical. On the
other hand, sinusoidal solutions to model (\ref{model compacton
bazeia}) have been derived in \cite{babichev}. However, no
compacton-like interpretation has been given}.
\\
The total energy of the single soliton state is
\begin{equation}
E=2\lambda^2a^2 \sqrt{\frac{\tilde{M}}{\lambda}}. \label{compacton
energy}
\end{equation}
Of course, as one expects, the scalar field reaches its vacuum
value at a finite distance. There is no exponential tail as
in the case of the $\phi^4$ kink. Thus, it is straightforward to
generalize it to a solution describing a chain of solitons and
anti-solitons with total topological charge $0$ or $\pm 1$.
Moreover, in this configuration each constituent soliton
(anti-soliton) does not interact with its neighbors. They just do
not "see" each other. Therefore, in such a chain solution the
positions of elementary solitons are arbitrary, provided that
after each soliton there appears an anti-soliton.
%%%%%%%%%%%%%%%%%%%%%%%%%%%%%%%%%%%%%%%%%%%%%%%%%%%%%%%%%%%%%%
\subsection{Linear stability}
%%%%%%%%%%%%%%%%%%%%%%%%%%%%%%%%%%%%%%%%%%%%%%%%%%%%%%%%%%%%%
In this section 
we shall demonstrate the linear stability of the
compactons (\ref{compacton sol}).
Here we closely follow the stability
analysis of Ref. \cite{bazeia}. We introduce general fluctuations around
a static (compacton) solution, $\xi (x,t) = \xi (x) + \eta (x,t)$ (here $\xi
(x)$ is the compacton solution, and $\eta (x,t)$ is the fluctuation field) and
insert this expression into the action of a general Lagrangian $L(v,\xi)$
(remember $v\equiv (1/2) \xi^\mu \xi_\mu$). The part of the action quadratic
in the fluctuation $\eta$, which is relevant for the stability analysis, is
\begin{equation}
S^{(2)} = \int d^2x \left( \frac{1}{2}L_v \eta^\mu \eta_\mu + L_{vv}
\frac{1}{2} (\xi^\mu \eta_\mu )^2 + L_{\xi \xi } \frac{1}{2} \eta^2 +
L_{\xi v}\eta \xi^\mu \eta_\mu \right)
\end{equation}
or, after using the identity
\begin{equation}
2L_{\xi v} \eta \xi^\mu \eta_\mu = \partial_\mu (L_{\xi v} \eta^2 \xi^\mu )
- \eta^2 \partial_\mu (L_{\xi v} \xi^\mu ) ,
\end{equation}
\begin{equation}
S^{(2)} = \frac{1}{2} \int d^2x \left( L_v \eta^\mu \eta_\mu + L_{vv}
(\xi^\mu \eta_\mu )^2 + L_{\xi \xi }  \eta^2 -
\partial_\mu (L_{\xi v} \xi^\mu ) \eta^2 \right) .
\end{equation}
The linear equation for the fluctuation field following from this action is
\begin{equation}
\partial_\mu \left( L_v \eta^\mu + L_{vv} \xi^\mu \xi_\alpha \eta^\alpha
\right) - [ L_{\xi \xi } -\partial_\mu (L_{\xi v} \xi^\mu )]\eta =0.
\end{equation}
Now we take into account that $\xi $ is static, and we replace $v$ by its
static limit $w\equiv -(1/2) \xi_x^2$. Further, we assume that
\begin{equation}
\eta (x,t)= \cos (\omega t)\eta (x) .
\end{equation}
The resulting linear ODE for $\eta (x)$ is
\begin{equation}
-\partial_x [(L_w +2L_{ww}w) \eta_x ] - [L_{\xi \xi} + \partial_x 
(L_{\xi w} \xi_x )] \eta = \omega^2 L_w \eta .
\end{equation}
For the specific class of Lagrangians $L= F(v) - U(\xi )$ this simplifies to
\begin{equation}
-\partial_x [(F_w +2 F_{ww} w)\eta_x ] + U_{\xi\xi} \eta =\omega^2 F_w \eta .
\end{equation}
Next, we specialize to the Lagrangian 
 (\ref{model compacton}) such that
\begin{equation}
F=4\tilde M |w| w \, ,\quad U= 3\lambda^2 (\xi^2 -a^2)^2
\end{equation}
and arrive at the equation
\begin{equation} \label{stab-eq}
-12 \tilde M^2 \partial_x (\xi_x^2 \eta_x ) + 12 \lambda^2 (3\xi^2 -a^2)\eta
= 4\tilde M^2 \omega^2 \xi_x^2 \eta .
\end{equation}
This expression must now be evaluated for the compacton solution
(\ref{compacton sol}) for $\xi (x)$.
In the outer region of the compacton,
i.e., in the region $|x| > \frac{\pi}{2} \sqrt{\frac{\tilde M}{\lambda}}$
where $\xi =\pm a = $const., obviously only the solution $\eta =0$ is
possible.  As we want $\eta$ to be continuous at the boundary of the
compacton, a general $\eta (x)$  should go to zero at the compacton
boundaries. The corresonding space of functions may 
be divided into an even and an odd 
subspace under the reflection $x\to -x$, and basis functions for the two
subspaces are
\begin{equation}
\eta_n (x) = \left\{
\begin{array}{lc}
0 & x \leq - \frac{\pi}{2} \sqrt{\frac{\tilde{M}}{\lambda}}  \\
 \cos (2n+1)\sqrt{\frac{\lambda}{\tilde{M}}}x &
-\frac{\pi}{2}\sqrt{\frac{\tilde{M}}{\lambda}} \leq x \leq
\frac{\pi}{2}\sqrt{\frac{\tilde{M}}{\lambda}}  \\
0 & x \geq \frac{\pi}{2}\sqrt{\frac{\tilde{M}}{\lambda}}
\end{array}
\right. \label{eta-x}
\end{equation}
for the even subspace (here $n=0, \ldots \infty$) and
\begin{equation} 
\zeta_m (x) = \left\{
\begin{array}{lc}
0 &  x \leq - \frac{\pi}{2} \sqrt{\frac{\tilde{M}}{\lambda}} \\
 \sin 2m \sqrt{\frac{\lambda}{\tilde{M}}}x &
-\frac{\pi}{2}\sqrt{\frac{\tilde{M}}{\lambda}} \leq x \leq
\frac{\pi}{2}\sqrt{\frac{\tilde{M}}{\lambda}} \\
0 &  x \geq \frac{\pi}{2}\sqrt{\frac{\tilde{M}}{\lambda}} 
\end{array}
\right. \label{zeta-x}
\end{equation}
for the odd subsapce  (here $m=1, \ldots ,\infty $).
The restriction to this class of functions will be important in the
stability analysis below.
Observe that the first derivative of $\eta$ is not continuous at the
boundary. This is consistent with the fact that the compacton itself is
continuous together with its first derivative. Also, Eq. (\ref{stab-eq}) is
well-defined everywhere, because $\eta_x$ is always multiplied by zero at the
points of discontinuity. \\
For linear stability, the eigenvalue $\omega^2$ at the r.h.s. of
Eq. (\ref{stab-eq}) has to be positive semi-definite, $\omega^2 \ge 0$. For
this to hold, the linear differential operator acting on $\eta$ at the
l.h.s. of Eq.  (\ref{stab-eq}) should be a positive semi-definite operator on
the space of functions (\ref{eta-x}), (\ref{zeta-x}). 
In order to demonstrate this, we rewrite
Eq. (\ref{stab-eq}) like
\begin{equation}
\tilde H \eta = 4\tilde M^2 \omega^2 \xi_x^2 \eta 
\end{equation}
where
\begin{eqnarray}
\tilde H &=& -12 a^2 \tilde M^2 \lambda \cos^2 
\sqrt{\frac{\lambda}{\tilde{M}}}x
\partial_x^2 + 24 a^2 \lambda^\frac{3}{2} \tilde M^\frac{1}{2}
\sin \sqrt{\frac{\lambda}{\tilde{M}}}x \cos \sqrt{\frac{\lambda}{\tilde{M}}}x
\partial_x \nonumber \\
&&+ 12 \lambda^2 a^2 (3 \sin^2 \sqrt{\frac{\lambda}{\tilde{M}}}x -1).
\end{eqnarray}
It is useful to introduce the new coordinate $y=\sqrt{\frac{\lambda}{\tilde M}}
  x $ and to rewrite
\begin{equation}
\tilde H = 12 a^2 \lambda^2 H
\end{equation}
with
\begin{equation}
H= -\cos^2 y \partial_y^2 +2 \sin y \cos y \partial_y +3\sin^2 y -1.
\end{equation}
We now want to demonstrate that the operator $H$ 
is positive semi-definite on the space of functions which are zero for
$|y|\ge \frac{\pi}{2}$ and continuous at the compacton boundaries
$y=\pm \frac{\pi}{2}$. This space may be divided into an even and an odd 
subspace under the reflection $y\to -y$, and these two subspaces may be 
treated separately, because the operator $H$ is even and does not mix
the two subspaces.  
A basis for the even subspace is (here $n=0, \ldots \infty$)
\begin{equation}
\eta_n (y) = \left\{
\begin{array}{lc}
0 & y \leq - \frac{\pi}{2}  \\
 \cos (2n+1)y &
-\frac{\pi}{2} \leq y \leq
\frac{\pi}{2}  \\
0 & y \geq \frac{\pi}{2} 
\end{array}
\right. \label{eta-y}
\end{equation}
whereas a basis for the odd subsapce is (here $m=1, \ldots ,\infty $)
\begin{equation} \label{odd-space}
\zeta_m (y) = \left\{
\begin{array}{lc}
0 & y \leq - \frac{\pi}{2}  \\
 \sin 2my &
-\frac{\pi}{2} \leq y \leq
\frac{\pi}{2}  \\
0 & y \geq \frac{\pi}{2} .
\end{array}
\right. \label{zeta-y}
\end{equation}

Next, we want to prove positive semi-definiteness of $H$ on the two subspaces.
For the even subspace we find
\begin{eqnarray}
&& \cos (2m+1)y H\cos (2n+1)y = 
\\ \nonumber &&
(n^2 +n+\frac{1}{2})[\cos 2(m-n)y + \cos
(2(m+n+1)y ] +
\\ \nonumber &&
\frac{1}{2} (n^2 -1)[\cos 2(m-n+1)y + \cos 2(m+n)y ] +
\\ \nonumber &&
\frac{1}{2} (n² +2n)[\cos 2(m-n-1)y + \cos 2(m+n+2)y ]
\end{eqnarray}
and, therefore,
\begin{eqnarray}
&& \langle m| H | n \rangle \equiv \int_{-\frac{\pi}{2}}^{\frac{\pi}{2}}
dy \cos (2m+1)y H\cos (2n+1)y = 
\\ \nonumber &&
\pi [(n^2 +n+\frac{1}{2})(\delta_{m,n} - \delta_{m,0} \delta_{n,0}) +
\\ \nonumber &&
\frac{1}{2} (n^2 -1) \delta_{m,n-1} + \frac{1}{2}(n^2 +2n)\delta_{m,n+1}]
\end{eqnarray}

We have to demonstrate positive semi-definiteness for a general vector
\begin{equation}
|v\rangle = \sum_{n=0}^\infty c_n \cos (2n+1)y
\end{equation}
where, however, we will restrict to normalizable vectors $v$. A normalizable
vector may always be approximated to arbitrary precision by a vector
\begin{equation}
|v\rangle = \sum_{n=0}^N c_n \cos (2n+1)y
\end{equation} 
for sufficiently large but finite $N$, 
therefore we will restrict to this case in the
sequel. 

Before continuing, we remark that 
the basis function $\cos y$ for $n=0$ is a zero mode of the operator $H$,
which is related to the translational invariance of the compactons, see Ref.
\cite{bazeia} for a more detailed discussion. 
Therefore, all matrix elements with $m=0$ or $n=0$ are
zero, and we may assume $c_0 =0$ without loss of generality. Taking this fact
into account, we find
\begin{eqnarray}
&& \langle v | H | v \rangle = \sum_{m,n=1}^N c_n \bar c_m \langle m|H|n\rangle
  =
\\ \nonumber &&
\pi \sum_{n=1}^N [c_n \bar c_n (n^2 +n+\frac{1}{2}) + c_n \bar c_{n-1}
\frac{1}{2}(n^2 -1) + c_n \bar c_{n+1} \frac{1}{2} (n^2 +2n)] \ge
\\ \nonumber &&
\pi \sum_{n=1}^N [c_n \bar c_n (n^2 +n+\frac{1}{2}) - |c_n | |\bar c_{n-1} |
\frac{1}{2}(n^2 -1) - |c_n | |\bar c_{n+1} | \frac{1}{2} (n^2 +2n)] =
\\ \nonumber &&
\pi \sum_{n=1}^N [|c_n|^2  (n^2 +n+\frac{1}{2}) - |c_n| | c_{n-1}|
(n^2 -1)]
\end{eqnarray}
where $c_{N+1}\equiv 0$ by assumption. We now want to prove that the above
expression is positive semi-definite. 
Positive semi-definiteness of this expression is implied by the inequality
\begin{equation} \label{fund-ineq}
\sum_{n=1}^N [|c_n|^2   - |c_n| | c_{n-1}|] \ge 0
\end{equation}
because of the inequality
\begin{equation}
 n^2 +n+\frac{1}{2} \ge n^2 -1 .
\end{equation}
Finally, inequality (\ref{fund-ineq}) may be proved easily with the help of
Hoelder's inequality.  

In fact, Hoelder's inequality reads
\begin{equation}
|\sum_{n=1}^N a_n b_n | \le \left( \sum _{n=1}^N |a_n|^p \right) ^\frac{1}{p}
\left( \sum _{n=1}^N |b_n|^q \right) ^\frac{1}{q} 
\end{equation}
where 
\begin{equation}
\frac{1}{p} + \frac{1}{q} =1.
\end{equation}
Now we set $p=q=2$ and $a_n =|c_n|$, $b_n = |c_{n-1}|$ and get
\begin{equation} \label{ineq-1}
\sum_{n=1}^N |c_n| |c_{n-1} | \le \left( \sum _{n=1}^N |c_n|^2 
\right) ^\frac{1}{2}
\left( \sum _{n=1}^N |c_{n-1}|^2 \right) ^\frac{1}{2}. 
\end{equation}
Further we have
\begin{equation}
\left( \sum _{n=1}^N |c_{n-1}|^2 \right) ^\frac{1}{2} =
\left( \sum _{n=0}^{N-1} |c_{n}|^2 \right) ^\frac{1}{2} \le
\left( \sum _{n=1}^N |c_{n}|^2 \right) ^\frac{1}{2} 
\end{equation}
where we used $c_0 =0$. Inserting this last inequality into (\ref{ineq-1})
just gives the inequality (\ref{fund-ineq}), which we wanted to prove.

The proof for the odd subspace (\ref{odd-space}) is completely analogous.
Indeed, we find
\begin{eqnarray}
&& \sin 2my H\sin 2ny = 
\\ \nonumber &&
(n^2 +\frac{1}{4})[\cos 2(m-n)y - \cos
(2(m+n)y ] +
\\ \nonumber &&
\frac{1}{2} (n^2 +n +\frac{3}{4})[\cos 2(m-n-1)y - \cos 2(m+n+1)y ] +
\\ \nonumber &&
\frac{1}{2} (n^2 -n + \frac{3}{4})[\cos 2(m-n+1)y - \cos 2(m+n-1)y ]
\end{eqnarray}
and, therefore,
\begin{eqnarray}
&& \langle m| H | n \rangle \equiv \int_{-\frac{\pi}{2}}^{\frac{\pi}{2}}
dy \sin 2my H\sin 2ny = 
\\ \nonumber &&
\pi [(n^2 +\frac{1}{4})\delta_{m,n}  +
\frac{1}{2} (n^2 +n +\frac{3}{4}) \delta_{m,n+1} + 
\frac{1}{2}(n^2 -n +\frac{3}{4})\delta_{m,n-1}]
\end{eqnarray}
For a general vector 
\begin{equation}
|v\rangle = \sum_{n=0}^N c_n \sin 2ny
\end{equation} 
we therefore get
\begin{eqnarray}
\label{ineq-2}
&& \langle v | H | v \rangle = \sum_{m,n=1}^N c_n \bar c_m \langle m|H|n\rangle
  =
\\ \nonumber &&
\pi \sum_{n=1}^N [c_n \bar c_n (n^2 +\frac{1}{4}) + c_n \bar c_{n-1}
\frac{1}{2}(n^2 -n + \frac{3}{4}) + c_n \bar c_{n+1} \frac{1}{2} (n^2 +n +
\frac{3}{4} )] \ge
\\ \nonumber &&
\pi \sum_{n=1}^N [c_n \bar c_n (n^2 +\frac{1}{4}) - |c_n | |\bar c_{n-1} |
\frac{1}{2}(n^2 -n +\frac{3}{4}) - |c_n | |\bar c_{n+1} | \frac{1}{2} 
(n^2 +n + \frac{3}{4})] =
\\ \nonumber &&
\pi \sum_{n=1}^N [|c_n|^2  (n^2 +\frac{1}{4}) - |c_n| | c_{n-1}|
(n^2 -n + \frac{3}{4})] .
\end{eqnarray}
Using the inequality
\begin{equation}
n^2 + \frac{1}{4} \ge n^2 -n + \frac{3}{4}
\end{equation}
positive semi-definiteness of expression (\ref{ineq-2})
is again implied by the
inequality (\ref{fund-ineq}), which has been proved above.

Finally, let us mention that alternative stability proofs, somewhat
different in spirit to the one presented here, have been given recently 
in \cite{det-brane} (see Section 6 of that reference), and in \cite{Baz1}.

%%%%%%%%%%%%%%%%%%%%%%%%%%%%%%%%%%%%%%%%%%%%%%%%%%%%%%%%%%%%%%
\subsection{Time dependence}
%%%%%%%%%%%%%%%%%%%%%%%%%%%%%%%%%%%%%%%%%%%%%%%%%%%%%%%%%%%%%
Even though the static Bogomolny solitons are identical in the
quartic $\phi^4$ model (both types (\ref{model compacton}),
(\ref{model compacton bazeia})) and in the model discussed by
Arod\'{z} et. al. their dynamics should be different,
as the Lagrangians have quite different structure. In
the standard case the dynamics is controlled by the wave equation
with a nonlinear (potential) term. It is unlike the quartic
$\phi^4$ model where a more complicated structure emerges.
\\ \\
It is not our aim to comprehensively discuss time dependent
configurations, but some interesting remarks can be easily made.
\\
In particular, let us consider the quartic $\phi^4$ model of the
type (\ref{model compacton bazeia}) previously discussed in
\cite{babichev}, \cite{bazeia}. The full equation of motion is
\begin{equation}
4\tilde{M}^2\partial_{\mu} \left[ \xi_{\nu}^2 \xi^{\mu} \right]
+U_{\xi}=0. \label{eom corr}
\end{equation}
Due to the symmetry between the time and space coordinates, one
may find the following solution
\begin{equation}
\xi (t) = \left\{
\begin{array}{lc}
- a & t \leq - \frac{\pi}{2} \sqrt{\frac{\tilde{M}}{\lambda}} \\
a \sin \sqrt{\frac{\lambda}{\tilde{M}}}t &
-\frac{\pi}{2}\sqrt{\frac{\tilde{M}}{\lambda}} \leq t \leq
\frac{\pi}{2}\sqrt{\frac{\tilde{M}}{\lambda}}  \\
a & t \geq \frac{\pi}{2}\sqrt{\frac{\tilde{M}}{\lambda}},
\label{compacton time sol}
\end{array}
\right.
\end{equation}
which represents a smooth transition
from one vacuum state to the other, collectively for all $x$.
The generalization to a solution
representing a composition of such collective jumps is
straightforward.
\\
Further, this solution is a
solution to the first order time-dependent differential equation
\begin{equation}
\xi^4_t=\frac{\lambda^2}{ \tilde{M}^2 }(\xi^2-a^2)^2 , \label{time
bogom}
\end{equation}
which can be viewed as a purely time dependent version of the
static Bogomolny equation. Similarly as in the usual case, the
solution of this Bogomolny equation gives a topology changing
configuration. Here, however, the two distinct vacua are connected
by a configuration depending on the time coordinate instead of the space
coordinate.
\\
As a consequence of (\ref{time bogom}), the corresponding energy
reads
\begin{equation}
E=\int_{-\infty}^{\infty} L_v \xi^2_t-L \;\; dx =
\int_{-\infty}^{\infty} -3 \tilde{M}^2
\xi^4_t+3\lambda^2(\xi^2-a^2)^2 \;\; dx =0.
\end{equation}
Thus, this solution is in fact a zero energy
solution.\footnote{One can observe that while the usual space Bogomolny
equation leads to vanishing $T^{11}$, its time-like version
enforce $T^{00}=0$ identically.} In other words, in this model
the global change of the vacuum costs nothing. It is really
remarkable that in the model with two separated vacua such a zero
energy transition is possible. This solution might suggest that we
have a kind of  dynamical restoration of the global $Z_2$
symmetry.
\\
Moreover, for a general time-dependent configuration the energy is
given by
\begin{equation}
E= \int_{-\infty}^{\infty} -3 \tilde{M}^2 \xi^4_t+2\tilde{M}^2
\xi^2_x\xi^2_t+\tilde{M}^2\xi^4_x+V \;\; dx.
\end{equation}
Therefore, this expression is not bounded from below and may take
arbitrarily large negative values. This might cause stability
problems in some applications of the model. In terms of general relativity
language, the model does not obey the null energy condition. 
It does, however, obey the hyperbolicity condition
\begin{equation} \label{hyperb}
1+2v\frac{L_{vv}}{L_v}>0
\end{equation}
for arbitrary (background) field configurations, and therefore the evolution
of small fluctuations will be hyperbolic for general backgrounds. 
In particular, the linear stability analysis of Section 4.4 remains
unchanged for this model and, therefore, small fluctuations around the 
compacton solutions behave the same way in both models.
A more detailed discussion of the roll of 
the null energy condition and the global
hyperbolicity condition in $k$-essence models can be found, e.g., 
in  Ref. \cite{BaMuVi}. 
\\ \\
The quartic $\phi^4$ model of the type (\ref{model compacton})
does not possess the zero-energy, topology changing solutions nor
has the problem of unbounded energy (of course, it also obeys the
hyperbolicity condition (\ref{hyperb})). 
On the contrary, the energy is positive
definite and reads
\begin{equation}
E=\int_{-\infty}^{\infty} 3 \tilde{M}^2 |\xi^2_t-\xi^2_x| \xi^2_t
+ \tilde{M}^2 |\xi^2_t-\xi^2_x| \xi^2_x +V \geq 0.
\end{equation}
Other possible modifications of Eq. (\ref{eom corr}) to cure the
stability problems mentioned above have been discussed in Ref.
\cite{babichev}.
%%%%%%%%%%%%%%%%%%%%%%%%%%%%%%%%%%%%%%%%%%%%%%%%%%%%%%%%%%%%%%
\subsection{Quartic sine-Gordon compactons}
%%%%%%%%%%%%%%%%%%%%%%%%%%%%%%%%%%%%%%%%%%%%%%%%%%%%%%%%%%%%%
Another example worth discussing is the quartic model with the
famous sine-Gordon potential \cite{dusual}
\begin{equation}
L=\tilde{M}^2 |\xi_{\mu}\xi^{\mu}|\xi_{\mu}\xi^{\mu} - \frac{3}{2}
\lambda^2 (1-\cos \xi). \label{model compacton sg}
\end{equation}
Thus, the Bogomolny equation is
\begin{equation}
\xi^4_x=\frac{\lambda^2}{ \tilde{M}^2 }(1-\cos \xi). \label{bogom
comp sg}
\end{equation}
It can be simplified to
\begin{equation}
\xi_x=\sqrt{\frac{\lambda}{ \tilde{M} }}\sqrt{\sin \frac{\xi}{2}}.
\label{comp sg1}
\end{equation}
After integration, one can find the simplest compacton solution
with  unit topological charge in the following exact form (see the Appendix
for a derivation)
\begin{equation}
\fl \xi (x) = \left\{
\begin{array}{lc}
0 & x \leq  x_1 \sqrt{\frac{\tilde{M}}{\lambda}}\\
4 \arctan \left[ \left( \frac{\mbox{sn\it{bx}} \;-(1+\sqrt{2})\;
\mbox{cn\it{bx}}}{  \mbox{sn\it{bx}}\; +(1+\sqrt{2}) \;
\mbox{cn\it{bx}}} \right)^2 \right] &
  x_1 \sqrt{\frac{\tilde{M}}{\lambda}} \leq x
 \leq x_2 \sqrt{\frac{\tilde{M}}{\lambda}}
 \\
2\pi & x \geq  x_2 \sqrt{\frac{\tilde{M}}{\lambda}}.
\end{array}
\right. \label{compacton sg sol}
\end{equation}
Here
$$ b=\frac{1}{4\sqrt{2}(2-\sqrt{2})} \sqrt{\frac{ \lambda}{\tilde{M}  }}$$
and $x_1, x_2$ are roots (the closest neighbors) of the algebraic
equations
$$ \mbox{sn\it{bx}}_1 \;-(1+\sqrt{2})\; \mbox{cn\it{bx}}_1=0, \;\;\;
\mbox{sn\it{bx}}_2 \;+(1+\sqrt{2})\; \mbox{cn\it{bx}}_2=0.$$
Moreover, $\mbox{sn}$ and $\mbox{cn}$ are the Jacobi elliptic
functions. Again, generalization to a multi-compacton
configuration is obvious.
%%%%%%%%%%%%%%%%%%%%%%%%%%%%%%%%%%%%%%%%%%%%%%%%%%%%%%%%%%%%%%
\subsection{Domain wall solutions in the full model}
%%%%%%%%%%%%%%%%%%%%%%%%%%%%%%%%%%%%%%%%%%%%%%%%%%%%%%%%%%%%%%
Finally, we consider the static soliton solutions of the full
model (\ref{model full domain}). The corresponding Bogomolny
equation is
\begin{equation}
3\tilde{M}^2 \xi_x^4+\frac{m^2}{2}
\xi^2_x=3\lambda^2(\xi^2-a^2)^2. \label{full domain eq mot 1}
\end{equation}
Although the equation is too complicated to derive an explicit
solution, we are able to describe its general properties.
\\
We begin the analysis with the calculation of the
asymptotic behavior of the scalar field near the vacuum values.
Assuming a smooth approach to the local minima of the
potential, that is, $\xi'_x \rightarrow 0$ for $x \rightarrow \pm
\infty$, we get that at sufficiently large distance from the
origin $\xi_x'^4 << \xi_x'^2$. Thus, the asymptotics is governed by
the quadratic term and the exponential tail should be visible for
all non-zero values of $m$. Specifically, for $\xi \rightarrow \pm
a$, one finds
\begin{equation}
\frac{m^2}{2} \xi^2_x=3\lambda^2(\xi^2-a^2)^2
\end{equation}
and
\begin{equation}
\xi (x) \simeq \pm a \left(1 -  2 e^{-\frac{2\sqrt{6} a\lambda}{m}
x} \right)
\end{equation}
for $x \rightarrow \pm \infty$. This shows that domain walls for the
full model are of the same type as the $\phi^4$ kink. In other
words, the exponential tail is a generic feature of topological
defects for this model. Therefore, the compacton solution is a
really critical case which happens for very special values of
the model parameters. In that sense, it is an isolated solution.
\\
While the asymptotics is entirely fixed by the quadratic term, the
behavior of the field in the vicinity of the center of the soliton
is influenced by the new quartic term. Indeed, one can find that
at the origin the solution is given by the following expression
\begin{equation}
\xi (x) \simeq A x, \;\;\; \mbox{if} \;\;\; x \simeq 0,
\end{equation}
where
$$A=\pm \frac{m}{\tilde{M} \sqrt{12}} \sqrt{\sqrt{1+
\left( \frac{12a\lambda \tilde{M}}{m^2}\right)^2}-1}.$$
Of course, in general one can find the domain wall solutions only
numerically. Then, from (\ref{full domain eq mot 1}) we
get the physically relevant root
\begin{equation}
\xi^2_x=\frac{m^2}{12 \tilde{M}^2} \left( \sqrt{1+\left( \frac{12
\tilde{M}\lambda}{m^2}\right)^2 (\xi^2-a^2)^2} -1\right).
\end{equation}
Numerical, finite solutions of this equation have been presented
in \cite{bazeia}. In fact, they are smooth functions monotonically
interpolating between two distinct vacua.
%%%%%%%%%%%%%%%%%%%%%%%%%%%%%%%%%%%%%%%%%%%%%%%%%%%%%%%%%%%%%%
\section{Half-compactons}
%%%%%%%%%%%%%%%%%%%%%%%%%%%%%%%%%%%%%%%%%%%%%%%%%%%%%%%%%%%%%%
So far we have investigated $k$-deformed models with compactons
where the potential term at a local minimum obeys two
requirements. Namely,
\begin{equation}
U'(\xi_0)=0, \;\;\; U''(\xi_0) \neq 0. \label{req halfcomp}
\end{equation}
However, if we allow for more general but still smooth potentials
then a new qualitatively different type of solitons can be
derived. Specifically, in the case of the quartic models we have
to assume that at least at one of the minima the first nonzero
derivative is the fourth one. Thus,
\begin{equation}
U'(\xi_0)= U''(\xi_0)=U^{(3)}(\xi_0)=0, \;\;\; U^{(4)}(\xi_0) \neq
0. \label{req comp}
\end{equation}
Of course, solutions of the type discussed here can exist in other
$k$-modified systems provided a suitable potential is introduced.
%%%%%%%%%%%%%%%%%%%%%%%%%%%%%%%%%%%%%%%%%%%%%%%%%%%%%%%%%%%%%%
\subsection{Models with half-compactons}
%%%%%%%%%%%%%%%%%%%%%%%%%%%%%%%%%%%%%%%%%%%%%%%%%%%%%%%%%%%%%%
The simplest case allowing for half-compactons is provided by
a potential with two minima
obeying conditions (\ref{req comp}) and (\ref{req halfcomp}),
respectively. The following quartic Lagrangian can serve as a
particular example
\begin{equation}
L=|\xi_{\mu}\xi^{\mu}|\xi_{\mu}\xi^{\mu} -
3\lambda^4(1-\xi)^2(1+\xi)^4. \label{half model 1}
\end{equation}
Then the static equation reads
\begin{equation}
\xi^4_x= \lambda^4(1-\xi)^2(1+\xi)^4
\end{equation}
and we find the solution
\begin{equation}
\xi=1-2\tanh^2 \frac{\lambda(x-x_0)}{\sqrt{2}}.
\end{equation}
This standard static configuration (note the exponential tails at
$x \rightarrow \pm \infty$) is well defined for all $x$ and
describes a topologically trivial solution joining the vacuum $\xi
=-1$ with itself. However, one can observe that it is a collection
of a {\it half-compacton} and an {\it anti half-compacton} glued
at $x=x_0$, where the second vacuum $\xi=1$ is reached.
\\
The half-compacton goes from the vacuum $\xi=1$ to $\xi=0$ and is
given by the formula
\begin{equation}
\xi=\left\{
\begin{array}{cc}
1 & x \leq x_0 \\
1-2\tanh^2 \frac{\lambda(x-x_0)}{\sqrt{2}}& x>x_0
\end{array}
\right.
\end{equation}
The construction of the anti half-compacton is
straightforward (it takes the vacuum value $\xi =1$ for $x\ge x_0$).
The name half-compacton is also obvious. It
denotes an object which reaches its two different
vacuum values in two different
ways. At one end of the defect (where the potential has a nonzero second
derivative) the approach is power-like (here parabolic). In other
words, at this end the soliton behaves like a compacton. On the
other hand, at the second end the exponential tail emerges and the
solution looks like a standard kink. In our example, the parabolic
approach occurs in the neighborhood of the vacuum at $\xi=1$,
\begin{equation}
\xi \simeq 1-\lambda^2 x^2, \;\;\; \mbox{for} \;\; x \simeq 0,
\end{equation}
whereas the vacuum at $\xi=-1$ is approached exponentially,
\begin{equation}
\xi \simeq -1+A e^{- \lambda \sqrt{2} x}, \;\;\; \mbox{for} \;\; x
\rightarrow \infty.
\end{equation}
The energy of the half-compacton configuration is
\begin{equation}
E=\frac{4^5}{2\sqrt{2}} \frac{16}{1155} \lambda^3.
\end{equation}
A slightly more complicated situation occurs if the quartic
Lagrangian possesses a potential with three vacua
\begin{equation}
L=|\xi_{\mu}\xi^{\mu}|\xi_{\mu}\xi^{\mu} - 3 \lambda^4
\xi^4(1-\xi^2)^2. \label{half model 2}
\end{equation}
Here, the minima at $\xi=\pm1$ are of type (\ref{req comp})
whereas the minimum at $\xi=0$ satisfies condition (\ref{req halfcomp}). One
can easily obtain the solution
\begin{equation}
\pm \ln \left|\frac{1+\sqrt{1-\xi^2}}{\xi} \right|=\lambda(x-x_0)
\end{equation}
or
\begin{equation}
\xi=\pm \frac{2 e^{\pm\lambda(x-x_0)}}{1+e^{\pm2\lambda(x-x_0)}}.
\end{equation}
This soliton configuration (with exponential tails) is well
defined for all $x$ and describes a solution joining the three
vacua $-1,0,+1$. Again, it is a collection of two half-compactons
glued at $x=x_0$.
\\
The first half-compacton joins the vacua $\xi=1$ and $\xi=0$
\begin{equation}
\xi=\left\{
\begin{array}{cc}
1 & x \leq x_0 \\
\frac{2 e^{-\lambda(x-x_0)}}{1+e^{-2\lambda(x-x_0)}}& x>x_0
\end{array}
\right.
\end{equation}
The second type of half-compacton connects $\xi=-1$ with $\xi=0$
\begin{equation}
\xi=\left\{
\begin{array}{cc}
-1 & x \leq x_0 \\
\frac{-2 e^{-\lambda(x-x_0)}}{1+e^{-2\lambda(x-x_0)}}& x>x_0
\end{array}
\right.
\end{equation}
Both half-compactons have the energy
\begin{equation}
E=\frac{2 }{35} \lambda^3.
\end{equation}
In the last example, we consider a model with infinitely many vacua
of both types  (\ref{req comp}) and (\ref{req halfcomp}),
\begin{equation}
L=|\xi_{\mu}\xi^{\mu}|\xi_{\mu}\xi^{\mu} - 3 (1-\cos \xi)(1+\cos
\xi)^2. \label{half model 3}
\end{equation}
The vacua with exponential approach are located at $\xi=(2k+1)\pi$, whereas
the vacua at $\xi=2k\pi$ are of the compacton type and are reached at
a finite distance. The static half-compacton solution
corresponding to the transition between vacua $2k\pi$ and
$(2k+1)\pi$ reads
\begin{equation}
\xi=\left\{
\begin{array}{cc}
0 & x \leq x_0 \\
\tilde{\xi} & x>x_0, \label{sol half 3}
\end{array}
\right.
\end{equation}
where $\tilde{\xi}$ is a solution to the equation
\begin{equation}
\arctan \sqrt{\left|\sin \frac{\xi}{2} \right|} +\arctan
\sqrt{\left|\sin \frac{\xi}{2} \right|} = \pm
\frac{\sqrt[4]{8}(x-x_0)}{2}.
\end{equation}
Of course, in Eq. (\ref{sol half 3}) the scalar field is
determined  modulo $2k\pi$ only.
%%%%%%%%%%%%%%%%%%%%%%%%%%%%%%%%%%%%%%%%%%%%%%%%%%%%%%%%%%%%%%
\subsection{Models with compactons and half-compactons}
%%%%%%%%%%%%%%%%%%%%%%%%%%%%%%%%%%%%%%%%%%%%%%%%%%%%%%%%%%%%%%
Finally, let us study a model with a four vacua potential
\begin{equation}
L=|\xi_{\mu}\xi^{\mu}|\xi_{\mu}\xi^{\mu} - 3 (1-\xi^2)^2(4-
\xi^2)^4. \label{comp half model}
\end{equation}
Here, the minima at $\xi=\pm 1$ are of the type (\ref{req halfcomp}) and the
minima at $\xi=\pm 2$ are of the type (\ref{req comp}). Therefore, one
might expect that this system allows for a compacton (joining $-1$
with $+1$) and two types of half-compactons. After some
calculation one finds the simplest compacton configuration
\begin{equation}
\xi=\left\{
\begin{array}{cc}
-1 & x-x_0 \leq - x_1 \\
\tilde{\xi} & -x_1<x-x_0<x_1 \\
1 & x-x_0 \geq x_1,
\end{array}
\right.
\end{equation}
where $\tilde{\xi}$ is a solution of the following expression for
$-1 \leq \tilde{\xi} \leq1$
\begin{equation}
\frac{\xi (\sqrt{1-\xi^2}-1)}{\xi^2+\sqrt{1-\xi^2}-1}=\pm
\frac{2}{\sqrt{3}} \tan 2\sqrt{3}(x-x_0). \label{sol half and
comp}
\end{equation}
and the joining point $x_1$ is
$$x_1=\frac{\pi}{4\sqrt{3}} .$$
Due to the complexity of Eq. (\ref{sol half and comp}) we are
not able to derive the explicit expression for the scalar field.
Nonetheless, the asymptotic behavior can be easily obtained
\begin{equation}
\xi \simeq \pm 1 \mp \frac{3^2}{2} (x\mp x_0 \mp
\frac{\pi}{4\sqrt{3}})^2, \;\;\; \mbox{for} \;\;\; x\mp x_0 \mp
\frac{\pi}{4\sqrt{3}} \simeq 0.
\end{equation}
The half-compactons are given by
\begin{equation}
\xi^+=\left\{
\begin{array}{cc}
1 & x-x_0 \leq   x_2\\
\tilde{\xi}^{+} & x-x_0>x_2,
\end{array}
\right.
\end{equation}
\begin{equation}
\xi^-=\left\{
\begin{array}{cc}
-1 & x-x_0 \leq x_2\\
\tilde{\xi}^{-} & x-x_0 >x_2,
\end{array}
\right.
\end{equation}
where $\tilde{\xi}^{\pm}$ are the positive and negative solutions of the
equation (here $ 1\leq \tilde{\xi}^{+}\leq 2$ and $ -2 \leq
\tilde{\xi}^{-} \leq -1$)
$$
\fl (7-4\sqrt{3})\ln|(\sqrt{\xi^2-1}+\xi)^2-(7-4\sqrt{3})|+ $$
\begin{equation}
(7+4\sqrt{3})\ln|(\sqrt{\xi^2-1}+\xi)^2-(7+4\sqrt{3})|=\pm
4\sqrt{3}(x-x_0)
\end{equation}
and $$x_2 = \frac{1}{4\sqrt{3}} \ln
\frac{(4\sqrt{3}+6)^{7+4\sqrt{3}}}{(4\sqrt{3}-6)^{7-4\sqrt{3}}}.$$
Here, once again one finds the parabolic approach to the vacua $\pm
1$. The exponential tail emerges if the scalar field tends to the
vacua $\pm 2$.
\\ \\
Let us make some general observations concerning topological
defects in the quartic Lagrangians. If two
nearest minima of the potential
are of the same kind we get either a compacton or
a standard
soliton. For potentials with $U''(\xi_0) \neq 0$ we get
compactons whereas for
$U'(\xi_0)=U''(\xi_0)=U^{(3)}(\xi_0)=0$ and $U^{(4)}(\xi_0) \neq
0$ we find kinks with an exponential tail. If the closest vacua
are of distinct types, then a half-compacton occurs.
\\
It is not difficult to convince oneself that half-compactons may be derived
also in models
with the usual, quadratic kinetic part. Then the corresponding
potential should have mixed $UV$ shape.
%%%%%%%%%%%%%%%%%%%%%%%%%%%%%%%%%%%%%%%%%%%%%%%%%%%%%%%%%%%%%%
\section{Conclusions}
%%%%%%%%%%%%%%%%%%%%%%%%%%%%%%%%%%%%%%%%%%%%%%%%%%%%%%%%%%%%%%
We think that the main achievement of the paper is the
demonstration of the close relation between the existence of
compactons and the appearance of a nonstandard kinetic term in the
Lagrangian. In fact, the compactons emerge as a result of the
specific relation between the kinetic and potential parts of the
Lagrangian. Thus, this kind of topological defects is no longer
one-to-one connected with $V$-shaped potentials. On the contrary,
they can be found for well behaved, analytical potentials if a
nonstandard kinetic term is allowed. Of course, this strongly
broadens the range of possible situations where such objets can be
relevant, indicating that the notion of compactons might have even
more interesting applications than expected previously.
\\
Moreover, we demonstrated that compactons do not necessarily
reach their vacuum value in the parabolic way. Depending on the
particular form of the kinetic term one can obtain arbitrary
power-like approaches. Similarly, the specific form of the
approximate scaling symmetry is also fixed by the mutual
relationship between the kinetic and potential parts of the action.
\\
As an example, we established the existence of
compactons as critical domain wall
solutions in a generalized Skyrme model.
Here the important point was that a model consisting of a real scalar field
$\xi$ with a quartic kinetic term and a regular potential (e.g., the quartic
$\phi^4$ model) could be embedded into the generalized Skyrme model,
therefore
compacton solutions will exist more generally for theories where such an
embedding is possible.
\\
We also established the existence of half-compactons, that is, of soliton
solutions which approach some vacua at a finite distance (like a compacton),
whereas other vacua are approached in the kink-like fashion (with an
exponential tail).
\\
Let us also mention that models with a
non-canonical kinetic term, which in principle can possess
compacton solutions, have been widely discussed in many contexts.
A class of models for which one can easily construct compacton
solutions is the following,
\begin{equation}
L=V(\xi) F(v). \label{string model}
\end{equation}
Such Lagrangians are considered as alternative models for a
canonical quintessence field. Moreover, this form of the
Lagrangian is suggested by string theory
\cite{kutasov}-\cite{hashimoto}. The pertinent static Bogomolny
equation reads
\begin{equation}
V(\xi)(F-2F_w w)=0
\end{equation}
and possesses solutions in the form $w=-a_i^2/2, \; a_i^2 \geq 0$
\cite{bazeia}. Thus
\begin{equation}
\xi = a_i x.
\end{equation}
This solution represents a compacton if the energy density is
localized in a finite region. Thus, we may assume that
\begin{equation}
V(\xi)=\left\{
\begin{array}{cc}
0 & \xi < - \xi_0 \\
V_0 & -\xi_0 \leq \xi \leq \xi_0 \\
0 & \xi > \xi_0
\end{array}
\right.
\end{equation}
Then, the corresponding energy
\begin{equation}
\fl E= -F(w=-a_i^2/2) \int_{-\infty}^{\infty} V(\xi) dx =
-\frac{F(w=-a_i^2/2)}{|a_i|} \int_{-\xi_0}^{\xi_0} V(\xi) d\xi
\end{equation}
is finite. For a detailed discussion of such a family of models see
\cite{bazeia}.
\\ \\
There are several directions in which the present work may be
continued. For example, dynamical properties of the half-compacton
can be analyzed. As they are asymmetric topological defects, their
interaction should depend on the mutual orientation of the ends.
It can be done at least in the simplest (and best understood) case
of the standard kinetic term and a $UV$-type potential. Another
issue worth investigating is a generalization of the inverse
scattering method (or other methods known from the standard
soliton theory) for $k$-deformed Lagrangians.
\\ \\
We believe that the established relations between compactons and
various $k$-defects (or more general fields with nonstandard
static gradient term) will open a new window for the application
of compactons.

\appendix
\section*{Appendix}
\setcounter{section}{1}

Here we calculate the quartic SG compacton. From (\ref{comp sg1}) we
get
\begin{equation}
2 \int \frac{dy}{\sqrt{\sin y}}=x-x_0,
\end{equation}
where $y=\xi/2$. After introducing $z^2=\tan (y/2)$ we find
\begin{equation}
\int \frac{dz}{\sqrt{1+z^4}}=\frac{1}{4\sqrt{2}} (x-x_0).
\end{equation}
This integral can be related to the elliptic integral of the
first type. Indeed, if we substitute
\begin{equation}
z=\frac{\tan \phi -(1+\sqrt{2})}{\tan \phi +(1+\sqrt{2})},
\label{z def}
\end{equation}
then we obtain
\begin{equation}
(2-\sqrt{2}) \int \frac{d\phi}{1-k^2 \sin^2
\phi}=\frac{1}{4\sqrt{2}} (x-x_0),
\end{equation}
where $$ k^2=\frac{4\sqrt{2}}{3+2\sqrt{2}}.$$ This expression can
be inverted
\begin{equation}
\phi = \mbox{am} \left[ \frac{1}{4\sqrt{2}(2-\sqrt{2})}
(x-x_0)\right].
\end{equation}
Now, inserting it into (\ref{z def}) and using $\sin \mbox{am}
\equiv \mbox{sn}$, $\cos \mbox{am} \equiv \mbox{cn}$ we arrive at
the result.

\ack A.W. gratefully acknowledges support from Adam
Krzy\.{z}anowski Fund and Jagiellonian University (grant WRBW
41/07). C.A. and J.S.-G. thank MCyT (Spain) and FEDER
(FPA2005-01963), and support from
 Xunta de Galicia (grant PGIDIT06PXIB296182PR and Conselleria de
Educacion). Further, C.A. acknowledges support from the Austrian
START award project FWF-Y-137-TEC and from the FWF project P161 05
NO 5 of N.J. Mauser. Finally, A.W. thanks H. Arod\'{z} and
P. Klimas for discussion.

\end{document}